\documentclass[prl,twocolumn,showpacs,preprintnumbers,amsmath,amssymb,aps]{revtex4-2}
\usepackage{graphicx}
\usepackage{amsmath}
\usepackage{amssymb}
\usepackage{epsfig}
\usepackage{amsthm}
\usepackage{bm}
\usepackage{color}
\def\be{\begin{equation}}	
\def\ee{\end{equation}}
\def\arr{\begin{array}{rll}}
\def\ea{\end{array}}
\def\bea{\begin{eqnarray}}
\def\eea{\end{eqnarray}}
\begin{document}
\title{Light absorption by weakly rough metal surfaces}
\author{Z. S. Gevorkian,$^{1,2}$ L. S. Petrosyan,$^{3}$ and T. V. Shahbazyan$^{3}$}
\affiliation{$^{1}$Alikhanyan National Laboratory, Alikhanian Brothers St. 2,  Yerevan 0036 Armenia
\\
$^{2}$Institute of Radiophysics and Electronics, Ashtarak-2 0203 Armenia 
\\
$^{3}$Department of Physics, Jackson State University, Jackson, Mississippi 39217 USA}
\begin{abstract}
We study light absorption by weakly rough metal surfaces with the roughness amplitude and correlation length smaller than the skin depth in metal. We develop a systematic perturbative approach for calculation of the absorptance in such systems and find that roughness-related absorptance variations are determined by an interplay between several system parameters which can result, in particular, in a greater absorption for smaller roughness amplitudes. We show that, for small-scale roughness, the absorptance variations are mainly caused by roughness-induced increase in effective volume of the surface layer, in which the incident light is predominantly absorbed. We argue that such  absorptance fluctuations between different samples, even though not related to any electron scattering processes, can appear as sample-to-sample variations of the Drude scattering rate reported in recent measurements of the metal dielectric function.
\end{abstract}

\maketitle

\section{Introduction}
Roughness is a property of materials that persists even for the best prepared samples.
Numerous papers are devoted to the scattering of electromagnetic waves from rough surfaces and its applications (see \cite{simon2010} for a recent review). It is well understood \cite{Brown85,McGurn87,McGurn96,Johnson99,Soubret01,Demir03,Navarette09} that, if the characteristic size of roughness parameters is comparable with the wavelength of incident light $\lambda$, the multiple scattering from surface imperfections can lead to a diffuse angular distribution of scattered light and to various other effects such as enhanced backscattering and weak localization \cite{Mar90,Tran94,Pak97,Mendez87,McGurn85}. In the opposite case of weak roughness, when the characteristic roughness size is much smaller than the wavelength, the diffuse component of scattered light is suppressed and incident light is predominantly specularly reflected from the metal surface.

While the reflection and scattering  from rough metal surfaces have been extensively studied, much less attention was paid to the effect of roughness on the absorption of electromagnetic waves in the metal \cite{Bergstrom08}. However, this is an  issue of considerable importance since even a weak roughness can substantially affect the measurements of the metal dielectric function, especially of its imaginary part \cite{yang15}. Precise knowledge of the dielectric function is crucial, e.g., for understanding the electronic structure of metals, chemical bonding, and optical properties\cite{dressel02,kuzmany09,kronig31,taft61,ehrenreich62,Lewis68}. The dielectric function also determines many important parameters in plasmonics \cite{stockman-review}, such as surface plasmon propagation length, plasmon radiation rate, and nonradiative losses \cite{Olmon12,Rioux14,zenin2022}.

Early works on absorption of electromagnetic waves in rough metals were focused on the absorption enhancement due to excitation of plasmon polaritons on rough surfaces \cite{Fedders68,Ritchie68,Crowell70,Elson71,Kretschmann69,Juranek70,maradudin1,raether88,zayats05}. The experimental paper \cite{springer04} confirmed  the dominance of plasmon polaritons for absorption in silver films  for relatively short wavelengths $\lambda<400$ nm and large-scale roughness with the root mean square (rms) amplitude $\delta$ and correlation length $a$ exceeding 100 nm. These findings were later supported by numerical calculations as well \cite{deghani21}. However, for a weak roughness case ($\delta<15$ nm), the experiment \cite{springer04} reported a discrepancy between the absorptance $A$ calculated as $A=1-R$, where $R$ is the measured total reflectance from the  rough metal surface, and the absorptance $A$ that is measured directly using photothermal deflection spectroscopy \cite{pds81}. Notably, no such discrepancy was reported for large-scale roughness characterized by a much stronger absorption. 

The purpose of this paper is to develop a consistent perturbative approach to light absorption by weakly rough metal surfaces characterized by   Gaussian random profile function $h$ with rms amplitude $\delta$  and correlation length $a$. In contrast to the light scattering problem, here the relevant lengthscale  that characterizes the incident light is the metal skin depth $d$, which,  in the frequency region we consider, is much smaller (by a factor of $20-50$) than the light wavelength $\lambda$. Accordingly,  all three length scales $\delta$, $a$ and $d$ are assumed to be much smaller than the wavelength $\lambda$, and so, the absorption is governed by three dimensionless parameters  $\delta/a$, $\delta/d$, and $a/d$. By "weakly rough" surface we imply that the first two parameters are small but the last one can be arbitrary, so that the perturbation expansion we employ is carried up to the order $\delta^{2}$ and the final expressions for the absorptance are evaluated numerically. As we show in this paper, the interplay between these parameters is highly nontrivial and, in particular, can lead to a non-monotonic absorptance dependence on the roughness amplitude $\delta$  reported in the experiment \cite{springer04}. We also obtain analytic expressions for small-scale roughness when the third parameter is small as well.

A serious limitation of the perturbative approach to such systems  is that a small variation of air-metal interface profile function $h$ results in abrupt and significant change in the  electric field distribution due to very large difference in magnitude between the air and metal permittivities \cite{carminati}. We address this issue by extending the boundary conditions for unperturbed fields to the actual surface profile, and show that the accurate treatment of this problem significantly reduces the first-order absorptance corrections, which otherwise are excessively large. Furthermore, we find that, for small-scale roughness ($a/d\ll 1$), the main roughness-related  effect comes from \textit{increase} in effective volume of the surface layer, in which the incident light is predominantly absorbed,  rather than from light scattering from surface imperfections. In this regime, the absorptance variation relative to smooth-surface absorptance is estimated as $\Delta A\propto \delta^{2}/a^{2}$, which can fluctuate substantially between different samples characterized by  small $\delta$ but larger spread of $a$. We argue that such absorptance fluctuations can appear as sample-to-sample variations of the Drude scattering rate, reported in recent measurements of the complex dielectric function \cite{yang15}, even though they are not related to any electron scattering processes in metals.

The paper is organized as follows. In Sec. \ref{sec-initial}, we set out our perturbation approach to the absorption by weakly rough metal surfaces. In Sec. \ref{sec-abs}, we calculate various contributions to the absorptance  and derive the asymptotic expressions for the case of small-scale roughness. In Sec. \ref{sec-num}, we discuss the results of our numerical calculations for silver films, and in Sec. \ref{sec-conc} we present our concluding remarks.

\section{Perturbation approach to light absorption by weakly rough surfaces}
\label{sec-initial}

\begin{figure}[tb]
\centering
\includegraphics[width=0.8\columnwidth]{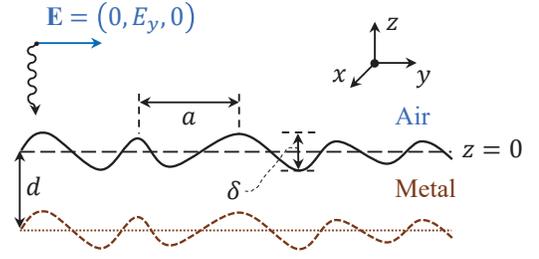}
\caption{Schematic view of the electromagnetic wave incident normally to  rough rough air-metal  interface. Dotted lines indicate the lower boundary of the surface layer, separated by skin depth $d$ from the interface (rough or smooth), in which light is predominantly absorbed.}
\label{fig1}
\vspace{-6mm}
\end{figure}

We consider a monochromatic electric field polarized along the $y$ axis that is incident normally upon the metal occupying  infinite region $z<h(y)$ (see Fig.~\ref{fig1}). We consider here the one-dimensional  roughness  case as it captures the essential features of weakly rough metals \cite{simon2010}, and assume that $h(y)$ is  a Gaussian random field with zero average, i.e., $\langle h(y)\rangle=0$, and correlation function
\begin{equation}
W(|y-y'|)=\langle h(y)h(y')\rangle=\delta^{2}e^{-(y-y')^2/2a^2}.
\end{equation}
Here, $\delta$ is the roughness rms amplitude  and $a$ is its correlation length, both of which are assumed to be much smaller than the incident light wavelength $\lambda$.

The fields above and below the air-metal interface are determined by the Maxwell equations
\begin{align}
&\bm{\nabla}\times {\bf E}=ik_{0}{\bf B},
\nonumber \\
&\bm{\nabla}^2{\bf E}-\bm{\nabla}(\bm{\nabla}\cdot{\bf E})+k_{0}^{2}\varepsilon({\bf r},\omega){\bf E}=0,
\label{max}
\end{align}
where $k_{0}=\omega/c$ is the wave vector in the air, $\omega$ is the wave frequency,  and the system dielectric permittivity $\varepsilon({\bf r},\omega)$ has the form
\begin{align}
\varepsilon({\bf r},\omega)
&=\Theta[z-h(y)]+\varepsilon(\omega)\Theta[h(y)-z]
\nonumber\\
&\approx\varepsilon_0(z,\omega)+\varepsilon_1({\bf r},\omega).
\label{dielcon}
\end{align}
Here, 
\begin{align}
\varepsilon_0(z,\omega)=\left\{\begin{array}{cc}
         1, \mbox{~~~~for~} z>0 \\
         \varepsilon(\omega), \mbox{for~} z<0
         \end{array}\right.
 \label{epszero}
 \end{align}
 is the permittivity for a smooth metal-air interface which, in the following, we refer to as the reference system, 
\begin{equation}
\varepsilon_1({\bf r},\omega)=[\varepsilon(\omega)-1]\delta(z)h(y),
\label{eps1}
\end{equation}
is the perturbation due to small variations of $h$, $\delta(z)$ is the Dirac delta function, and $\varepsilon(\omega)=\varepsilon'(\omega)+i\varepsilon''(\omega)$ is the  metal dielectric function.

We decompose the electric field as ${\bf E}={\bf E}_0+{\bf E}_s$, where ${\bf E}_0$ is the reference field and ${\bf E}_s$ is the scattered field due to surface roughness. These satisfy the equations (suppressing, for brevity, the $\omega$ dependence)
 \begin{align}
\nabla^2{\bf E}_0-\bm{\nabla}(\bm{\nabla}\cdot{\bf E}_0)+k_{0}^{2}\varepsilon_0(z){\bf E}_0=0,
\label{reffield}
\end{align}
and
\begin{align}
\nabla^2{\bf E}_s-\bm{\nabla}(\bm{\nabla}\cdot {\bf E}_s)+k_{0}^{2}\varepsilon({\bf r}){\bf E}_s=-k_{0}^{2}\varepsilon_1({\bf r}){\bf E}_0.
\label{scattfield}
\end{align}
The scattered fields can be obtained in a standard manner using the dyadic  Green's functions defined as \cite{maradudin1,maradudin2},
\begin{align}
\left[k_{0}^{2}\varepsilon_0(z)-\bm{\nabla}\bm{\nabla}+\nabla^2+k_{0}^{2}\varepsilon_1({\bf r})\right]\textbf{D}({\bf r},{\bf r}')=4\pi\delta({\bf r}-{\bf r}'),
\label{green}
\end{align}
where perturbation expansion over $\varepsilon_1$ is implied.

The reference field ${\bf E}_0$ can be chosen as the sum of  incident and reflected plane waves in the air and a transmitted plane wave in the metal for a system with a smooth air-metal interface. However,  this is not a good choice for the lowest order of perturbation expansion because, due to  large value of the metal permittivity for optical and infrared frequencies ($|\varepsilon'|\gg 1$), even a small change in the air-metal interface  position leads to abrupt and significant field variations \cite{carminati}. To avoid this issue, in the case of weak roughness, we can modify the reference field by extending it up to the actual interface: 
\begin{align}
\tilde{E}_{0y}({\bf r})=\left\{\begin{array}{cc}e^{-ik_{0}z}-re^{ik_{0}z},\quad \mbox{~for~} z>h(y),
\\
te^{ik_{-}z}, \quad \mbox{\qquad\qquad for~} z<h(y),
\end{array}\right.
\label{back}
\end{align}
where  $k_{-}=-k_{0}\sqrt{\varepsilon}$ (the negative sign ensures that the transmitted wave decays into metal). Note that $\tilde{E}_{0x}=\tilde{E}_{0z}=0$ for our choice of polarization. Here, the incident wave amplitude $E_{\rm inc}$ is taken to be unity, while $r$ and $t$ are the standard Fresnel coefficients of reflection and transmission for normal incidence:
\begin{equation}
r=\frac{\sqrt{\varepsilon}-1}{\sqrt{\varepsilon}+1}, \quad
t=\frac{2}{\sqrt{\varepsilon}+1}.
\label{modt}
\end{equation}
Accordingly, the field decomposition now has the form ${\bf E}=\tilde{\bf E}_0+\tilde{\bf E}_s$, where the modified scattered field  is expressed through the dyadic Green's function $\textbf{D}({\bf r},{\bf r}')$ as
\begin{equation}
\tilde{\textbf{E}}_{s}({\bf r})=-\frac{k_{0}^{2}}{4\pi}\int d{\bf r}'\textbf{D}({\bf r},{\bf r}')\tilde{\textbf{E}}_{0}({\bf r}')\tilde{\varepsilon}_1({\bf r}').
\label{scatf}
\end{equation}
Here, $\tilde{\varepsilon}_1({\bf r})=(\varepsilon-1)\delta[z-h(y)]h(y)$ is the modified perturbation obtained from Eq.~(\ref{eps1}) by the replacement $z\rightarrow z-h(y)$ in the $\delta$-function in order to make it consistent with the extended boundary conditions  (\ref{back}). Note that, while $\tilde{\varepsilon}_1({\bf r})$ and $\varepsilon_1({\bf r})$ coincide in the first order, the accurate choice of reference field leads to a significant reduction of excessively large first-order correction to the absorptance, as we show later in this paper.

For a monochromatic wave with frequency $\omega$, the power absorbed in a metal is given by  \cite{landau}
\begin{equation}
Q=\frac{\omega}{8\pi}\varepsilon''\int dV|{\bf E}|^2,
\label{loss}
\end{equation}
where integration is carried out over the metal volume. The absorptance $A$ is obtained by normalizing $Q$ by the incident energy flux $Q_{\rm inc}=c|E_{\rm inc}|^{2}S_{0}/8\pi$, where $S_{0}=L_{x}L_{y}$ is the normalization area. Using the above field decomposition, the absorptance averaged over the roughness configurations takes the form
\begin{equation}
A= \left \langle \frac{\varepsilon''k_{0}}{S_{0}}\int\! dV\! \left[|\tilde{{\bf E}}_0|^2+2\text{Re}(\tilde{{\bf E}}_0^*\cdot\tilde{\bf E}_s)+|\tilde{\bf E}_s|^2\right]\right \rangle.
\label{bulk2}
\end{equation}
Below we evaluate all contributions to Eq.~(\ref{bulk2}) perturbatively, i.e., up to the order $\delta^{2}$. Specifically, we assume that the dimensionless parameters  $\delta/\lambda$, $\delta/d$ and $\delta/a$ are small, but no restriction is imposed on the parameter $a/d$ in the general expressions derived in the next section. In addition, we present below the analytical expressions in the \textit{asymptotic} regime  $a/d\ll 1$, which are valid for the metal films with small-scale roughness.

\section{Calculation of the absorptance}
\label{sec-abs}

\subsection{Reference field contribution}
\label{sec-ref}

We start with the first term in Eq.~(\ref{bulk2}) describing the reference field contribution:
\begin{equation}
A_{r}=\frac{\varepsilon''k_{0}}{S_{0}}\left\langle\int dV |\tilde{{\bf E}}_0|^2\right\rangle.
\label{bulbac}
\end{equation}
The integration over the metal volume can be decomposed as $\int dV=\int dz \int dS$, where $dS=dxdy\sqrt{1+h'^2}$ is the differential area of the  surface with a $z=h(y)$ profile. Taking into account the extended boundary conditions  Eq.~(\ref{back}) and using Eq.~(\ref{modt}), we have 
\begin{align}
A_{r}=\frac{\varepsilon''k_{0}|t|^{2}}{S_{0}}
\left\langle\int dxdy\sqrt{1+h'^2(y)}\int_{-\infty}^hdze^{-2\kappa k_{0}z}\right\rangle,
\label{bulbac2}
\end{align}
where we  adopted the standard notation $\sqrt{\varepsilon}=n+i\kappa$ for the complex refraction index. Integrating over $z$ and expanding the integrand over  $h$ and $h'$, we obtain
\begin{align}
A_{r}=\frac{A_{0}}{S_{0}}
\!\int\! dxdy\left[1+\frac{\langle h'^2(y)\rangle}{2}+\frac{2\langle h^2(y)\rangle}{d^{2}}\right],
\label{bulbac3}
\end{align}
where $A_{0}=\varepsilon''|t|^{2}/2\kappa$ is the absorptance for a smooth metal surface and $d=(k_{0}\kappa)^{-1}$ is  skin depth in the metal. Averaging over roughness configurations as $\langle h^2(y)\rangle=\delta^{2}$ and $\langle h'^2(y)\rangle=\delta^{2}/a^{2}$, we finally obtain
\begin{equation}
A_{r}=A_{0}\left (1+\frac{\delta^2}{2a^2}+\frac{2\delta^2}{d^2}\right ).
\label{bulkbackfin}
\end{equation}
%
In the case of small-scale roughness, $a\ll d$, the last term is small, and the absorptance variations  are determined solely by the roughness parameters: $\Delta A_{r}/A_{0}\approx \delta^{2}/2a^{2}$. This contribution originates from roughness-induced increase in effective volume of the region, in which the incident light is predominantly absorbed. This region can be visualized as a surface layer of thickness $d$ that is measured from the actual surface profile, as opposed to a layer of the same thickness but with smooth boundaries (see Fig.~\ref{fig1}).

\subsection{Interference term contribution}
\label{sec-int}

Consider now the second term in Eq.~(\ref{bulk2}) describing interference between the reference  and scattered fields,
\begin{equation}
A_{i}=  \frac{\varepsilon''k_{0}}{S_{0}} 2\text{Re}\left \langle\int\! dV \tilde{{\bf E}}_0^*\cdot\tilde{\bf E}_s\right \rangle,
\label{intterm}
\end{equation}
where $\tilde{\bf E}_s$ is given by Eq.~(\ref{scatf}). Up to the order $h^{2}$, we can present the scattered field as a sum of two terms, $\tilde{\bf E}_s=\tilde{\bf E}_s^{(1)}+\tilde{\bf E}_s^{(2)}$, corresponding, respectively, to the lowest and first order perturbation expansion of the Green function $\textbf{D}$ in Eq.~(\ref{scatf}). Accordingly, this contribution to the absorptance can also be split as $A_{i}=A_{i1}+A_{i2}$.

We start with the first contribution obtained by inserting into Eq.~(\ref{scatf}) the unperturbed Green's function $\textbf{D}_{0}$ obtained by setting $\varepsilon_{1}=0$ in Eq.~(\ref{green}). One might be tempted to think that since $\tilde{\bf E}_s^{(1)}\sim h$, the corresponding absorptance $A_{i1}$ would vanish after performing averaging over the roughness configurations. However, as we show below, the extended boundary conditions Eq.~(\ref{back}) for the modified reference field  $\tilde{{\bf E}}_0$ lead to a \textit{negative} contribution to the absorptance which balances out the excessive absorption increase due to scattered field penetration into the metal. 

To evaluate the average $C_{1}=\left \langle\int\! dV \tilde{{\bf E}}_0^*\cdot\tilde{\bf E}_s^{(1)}\right \rangle$, we introduce two-dimensional Fourier transform of the unperturbed Green's function as
\begin{equation}
\textbf{D}_{0}(\bm{\rho}-\bm{\rho}',z,z^{\prime})=\int\frac{d {\bf q}}{(2\pi)^2}\textbf{d}({\bf q},z,z')e^{i{\bf q}\cdot (\bm{\rho}-\bm{\rho}')},
\label{fourier}
\end{equation}
where $\bm{\rho}$ is a two-dimensional position vector in the $xy$ plane and $\textbf{d}({\bf q},z,z')$ is a matrix function in coordinate space (see below). Employing this expression in  Eq.~(\ref{scatf}), we present $C_{1}$ in the form
\begin{align}
C_{1}=-&\frac{k_{0}^{2}|t|^{2}}{4\pi}
(\varepsilon-1)  \int d\bm{\rho}d\bm{\rho}'
 \!\int\frac{d\bm{q}}{(2\pi)^2}e^{i\bm{q}\cdot(\bm{\rho}-\bm{\rho}')}
\nonumber\\
&\times
\left\langle \int_{-\infty}^h \!dz d_{yy}[\bm{q},z,h(y')]e^{-ik_{-}^*z}h(y')\right\rangle.
\label{intterm2}
\end{align}
The matrix function $\textbf{d}({\bf q},z,z')$ can be presented as  \cite{maradudin1} $\textbf{d}({\bf q},z,z^{\prime})=\textbf{S}^{-1}\textbf{g}(q,z,z^{\prime})\textbf{S}$, where the matrix function $\textbf{g}(q, z,z^{\prime})$ is tabulated in \cite{maradudin1,maradudin2}, while $3\times 3$ matrix $\textbf{S}$ has the following elements $S_{xx}=S_{yy}=q_x/q$, $S_{zz}=1$, $S_{xy}=-S_{yx}=q_y/q$, and $S_{xz}=S_{zx}=S_{yz}=S_{zy}=0$. For normal incidence and one-dimensional roughness profile function $h(y)$, the $\bm{\rho}$-integration in Eq.~(\ref{intterm2}) sets $q_{x}=0$, and we obtain
\begin{equation}
d_{yy}({\bf q},z,z')=-\frac{2\pi i k_1}{\varepsilon k_{0}^{2}}\left[\frac{k_1+\varepsilon k}{k_1-\varepsilon k}e^{ik_1(z+z')}-e^{-ik_1|z-z'|}\right],
\label{gxxyy}
\end{equation}
where the points $z$ and $z'$ are assumed inside the metal, and
\begin{align}
&k(q)=\left\{\begin{array}{cc}
(k_{0}^{2}-q^2)^{1/2},\quad q<\omega/c \nonumber \\
i(q^2-k_{0}^{2})^{1/2},\quad q>\omega/c \end{array}\right.\nonumber\\
&k_1(q)=-\left(\varepsilon k_{0}^{2}-q^2\right)^{1/2}.
\label{kandk}
\end{align}
Inserting Eq.~(\ref{gxxyy}) into Eq.~(\ref{intterm2}) and integrating over $z$, we obtain
\begin{align}
&C_{1}
=-\frac{|t|^{2}}{2\varepsilon}(\varepsilon-1)\int d\bm{\rho}\int\frac{dq}{2\pi}
\left\langle\frac{k_{1}}{k_{-}^*-k_1}e^{i(k_1-k_{-}^*)h(y)}\right. 
\nonumber\\
&\times
\left.\int dy'\left[\frac{k_1+\varepsilon k}{k_1-\varepsilon k}e^{ik_1h(y')}-e^{-ik_1h(y')}\right]
e^{iq(y-y')}h(y')\right\rangle,
\label{intterm3}
\end{align}
where $q\equiv q_y$. Expanding Eq.~(\ref{intterm3}) over $h$, performing averaging, over roughness configurations, and then calculating the integrals, the result can be presented in the form $C_1=C_1^{(1)}+C_1^{(2)}$, where
\begin{align}
C_1^{(1)}=-|t|^{2}(\varepsilon-1)S_{0}
\int\frac{dq}{2\pi}\frac{k(q)k_1(q)W(q)}{i[k_1(q)-\varepsilon k(q)]},
\label{interm4}
\end{align}
and
\begin{align}
C_1^{(2)}=-\frac{i|t|^{2}S_{0}}{\varepsilon}\,
\frac{(\varepsilon-1)k_1^3(q=0)W(y=0)}{[k_{-}^*-k_1(0)][k_1(0)-\varepsilon k(0)]}.
\label{interm5}
\end{align}
Here, $W(q)=\delta^2 a\sqrt{2\pi}e^{-q^2a^2/2}$ is the Fourier transform of correlation function $W(y)$ and $W(y=0)=\delta^2$. Note that $C_1^{(1)}$ originates from the averaging of two profile functions at different points, $W(y)=\langle h(y)h(y')\rangle$, when the first exponent in Eq.~(\ref{intterm3}) is expanded up to the first order in $h(y)$, while $C_1^{(2)}$ involves  averaging of two profile functions at coinciding points, $W(y=0)=\langle h(y)h(y)\rangle$, when the exponents in the square brackets are expanded.

Let us first estimate $C_1^{(2)}$. Substituting $k(0)$ and $k_1(0)$ from Eq.~(\ref{kandk}) into Eq.~(\ref{interm5}), we find
 \begin{equation}
C_1^{(2)}=-\frac{|t|^{2}\delta^{2}k_{0}}{2\kappa}\,S_{0}
\frac{\varepsilon-1}{\sqrt{\varepsilon}+1}.
 \label{qi2}
 \end{equation}
 It is easy to check that, in the optical and infrared spectral domain (i.e., $|\varepsilon'|\gg 1$), the corresponding  absorptance contribution is $A_{i1}^{(2)}=(\epsilon''k_{0}/S_{0})2\text{Re} C_1^{(2)} \approx 2A_{0}\delta^{2}k_{0}^{2}$, and, hence, it is suppressed by the small factor $|\varepsilon'|^{-1}$ as compared to the last term in Eq.~(\ref{bulkbackfin}). Therefore, this contribution can be neglected.

Turning to $C_1^{(1)} $, we introduce the dimensionless variable $x=qa$ in the integral and write
\begin{equation}
C_1^{(1)}=-\frac{2|t|^{2}\delta^{2}}{a}\,S_{0}(\varepsilon-1)I(\beta),
\end{equation}
where
\begin{align}
I(\beta)=\int_{0}^{\infty}\frac{dx}{i\sqrt{2\pi}}\frac{\sqrt{(\beta^2-x^2)(\varepsilon\beta^2-x^2)}
e^{-x^2/2}}
{\sqrt{\varepsilon\beta^2-x^2}+\varepsilon\sqrt{\beta^2-x^2}}
\label{intinter}
\end{align}
and  $\beta=k_{0}a$. In this way, we obtain the first interference contribution to the absorptance as
%
\begin{equation}
A_{i1}=\frac{2\epsilon''k_{0}}{S_{0}}\,
\text{Re}\, C_1
=-A_{0}\frac{8\delta^{2}}{da}\,\text{Re} \left [(\varepsilon-1)I(\beta)\right ].
\label{inttermfin}
\end{equation}
In the case of small-scale roughness $a\ll d$, the integral can be evaluated at  $\beta=0$, yielding $I(0)=1/\sqrt{2\pi}(\varepsilon+1)$. Finally, in the frequency domain $|\varepsilon'|\gg 1$, we obtain the asymptotic expression
\begin{equation}
A_{i1}\approx -A_{0}\frac{8\delta^{2}}{\sqrt{2\pi}da},
\label{interasymp}
\end{equation}
which has  a negative sign.

Turning now to the second interference contribution $A_{i2}$, we expand the Green's function $\textbf{D}$, defined by Eq.~(\ref{green}),  to the first order in $h$, and  present the second-order scattered field in Eq.~(\ref{scatf}) as 
\begin{align}
\tilde{\bf E}_s^{(2)}=\left (\frac{k_{0}^2}{4\pi}\right )^{2}\int & d{\bf r}'d{\bf r}''
\textbf{D}_{0}({\bf r},{\bf r}')\varepsilon_1({\bf r}')
\nonumber\\
&\times \textbf{D}_{0}({\bf r}',{\bf r}'')
\varepsilon_1({\bf r}'')\textbf{E}_{0}({\bf r}''),
\label{seconord}
\end{align}
where, in the order $h^{2}$, the modified reference field and perturbation can be replaced by the original ones. Evaluating $A_{i2}$ in a similar manner, we obtain 
\begin{equation}
A_{i2}=A_{0}\frac{4\delta^2}{ad}\,\text{Re}\left [\frac{(\varepsilon-1)^2I(\beta)}{i\kappa (\sqrt{\varepsilon}+1)}\right ].
\label{seccont}
\end{equation}
For small-scale roughness case $a\ll d$, we can use $I(0)=1/\sqrt{2\pi}(\varepsilon+1)$, and then, for $|\varepsilon'|\gg 1$, we obtain the asymptotic expression for this contribution as
\begin{equation}
A_{i2}\approx A_{0}\frac{4\delta^{2}}{\sqrt{2\pi}da}.
\label{ascoh}
\end{equation}
Finally, the full interference contribution $A_{i}=A_{i1}+A_{i2}$ is still negative and, for small-scale roughness, has the asymptotic form,
\begin{equation}
A_{i}\approx -A_{0}\frac{4\delta^{2}}{\sqrt{2\pi}da}.
\label{asymint}
\end{equation}

\subsection{Scattering term contribution}
\label{sec-scatt}

Consider now the scattered field contribution to the absorptance which we present in the form
\begin{equation}
A_{s}
= \frac{\varepsilon''k_{0}}{S_{0}} \left \langle \int\! dV \left (|\tilde{E}_{sx}|^2+|\tilde{E}_{sy}|^2+|\tilde{E}_{sz}|^2\right ) \right \rangle.
\label{bulkscat}
\end{equation}
After evaluating each term in the way outlined in the previous section, the result can be presented as 
\begin{equation}
A_{s}=A_{0}\frac{2\delta^2}{ad}|\varepsilon-1|^2 \left [I_{x}(\beta) +I_{y}(\beta) +I_{z}(\beta) \right ],
\label{bsxy}
\end{equation}
where
\begin{align}
I_{x}(\beta)=
\int_{0}^{\infty}\frac{dx}{\sqrt{2\pi}}
&\frac{|\beta^2-x^2|}{|\text{Im}\sqrt{\varepsilon\beta^2-x^2}|}
\nonumber\\
&\times\frac{x^2e^{-x^2/2}}{|\sqrt{\varepsilon\beta^2-x^2}+\varepsilon\sqrt{\beta^2-x^2}|^2},
\label{isx}
\end{align}
\begin{align}
I_{y}(\beta)=
\int_{0}^{\infty}\frac{dx}{\sqrt{2\pi}}
&\frac{|\beta^2-x^2|}{|\text{Im}\sqrt{\varepsilon\beta^2-x^2}|}
\nonumber\\
&\times\frac{|\varepsilon\beta^2-x^2|e^{-x^2/2}}{|\sqrt{\varepsilon\beta^2-x^2}+\varepsilon\sqrt{\beta^2-x^2}|^2},
\label{isy}
\end{align}
and $I_{z}(\beta)=I_{x}(\beta)$. In the small-scale roughness case $a\ll d$, using $I_{x}(0)=I_{y}(0)=I_{z}(0)=1/\sqrt{2\pi}|\varepsilon+1|^{2}$, we obtain  the asymptotic ($|\varepsilon'|\gg 1$) expression,
\begin{equation}
A_{s}\approx A_{0}\frac{6\delta^{2}}{\sqrt{2\pi}da}.
\label{asymscat}
\end{equation}
Finally, the full absorptance is obtained by summing up all contributions: $A=A_{r}+A_{i}+A_{s}$. The small-scale roughness asymptotic expression is obtained by collecting the corresponding terms from Eqs.  (\ref{bulkbackfin}), (\ref{asymint}) and (\ref{asymscat}):
\begin{equation}
A \approx A_{0}\left (1+\frac{\delta^{2}}{2a^{2}}+\frac{2\delta^{2}}{\sqrt{2\pi}da}+\frac{2\delta^{2}}{d^{2}}\right ).
\label{finabs}
\end{equation}
Concluding this section, we note that the accurate choice of reference field (\ref{back}) ensures the small magnitude of  first-order correction to the absorptance in the weak-roughness case. Specifically, had we chosen the standard, rather than extended, boundary conditions for reference fields, the third term in Eq.~(\ref{finabs}) would have increased \textit{fivefold}, signaling a poor choice of  basis set for the perturbation expansion. Although  the overall absorptance increases, relative to  smooth-surface absorptance, by the amount  $\propto \delta^{2}$, its precise behavior is determined by the interplay between two scales -- the skin depth $d$ and correlation length $a$, as we discuss in the next section.
Finally, note that general expressions  Eqs. (\ref{bulkbackfin}), (\ref{inttermfin}), (\ref{seccont}) and (\ref{bsxy}), used in the numerical calculations below,  are accurate up to the order $\delta^{2}$ with no other conditions, and therefore can be used for dielectric materials as well.  However, the analytical expression Eq.~(\ref{finabs}), obtained in the limit $a\ll d$, is only accurate for metals in the frequency region  $|\varepsilon'|\gg 1$.

%
\begin{figure}[tb]
\begin{center}
\includegraphics[width=0.9\columnwidth]{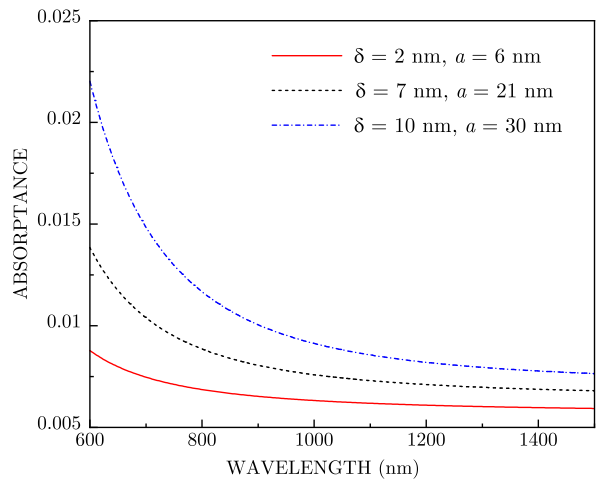}
\caption{\label{fig2} Calculated absorptance at several values of roughness amplitude $\delta$ plotted against incident light wavelength  for fixed ratio $\delta/a=1/3$.}
\end{center}
\vspace{-8mm}
\end{figure}
%

\section{Numerical results and discussion}
\label{sec-num}

Below we present the results of numerical calculations of absorptance for weakly rough opaque silver films. The roughness parameters were chosen in the range $\delta\ll d$ and $a\lesssim d$, while the experimental dielectric function of silver was used in all calculations \cite{christy72,Rioux14}. The wavelength interval is chosen from $\lambda=600$ to 1500 nm in order to avoid the influence of surface plasmon ($\lambda\approx350$ nm in silver) and of interband transitions, both of which lead to enhanced absorption not directly related to roughness. In this interval, the skin depth in silver is $d\approx 25$ nm  and   weakly depends on the wavelength \cite{yang15}. All numerical calculations were carried out using the full expression for absorptance $A$, while the small-scale roughness asymptotic expression (\ref{finabs}) is used to discuss qualitative features of obtained results.

In Fig.~\ref{fig2}, we show the calculated absorptance for several values of $\delta$ and $a$ at a fixed ratio $\delta/a=1/3$. In this case, the second term in Eq.~(\ref{finabs}) is unchanged for all curves. The overall scale of calculated absorptance is consistent with the results reported in the experiment \cite{springer04}, indicating that, in this frequency range, about 99\% of incident light is reflected back. We note that for the lowest curve ($a=6$ nm), the absorptance is nearly constant for $\lambda > 1000$ nm, consistent with weak frequency dependence of smooth-surface absorptance in the Drude regime: $A_{0}\sim \gamma/\omega_{p}$, where $\gamma$ and $\omega_{p}$ are the Drude scattering rate and plasma frequency, respectively (see below). At the same time, the curves calculated for larger values of correlation length $a$ exhibit enhanced absorptance at shorter wavelength due to roughness-assisted  excitation surface plasmon polaritons \cite{simon2010}.

%
\begin{figure}[tb]
\begin{center}
\includegraphics[width=0.9\columnwidth]{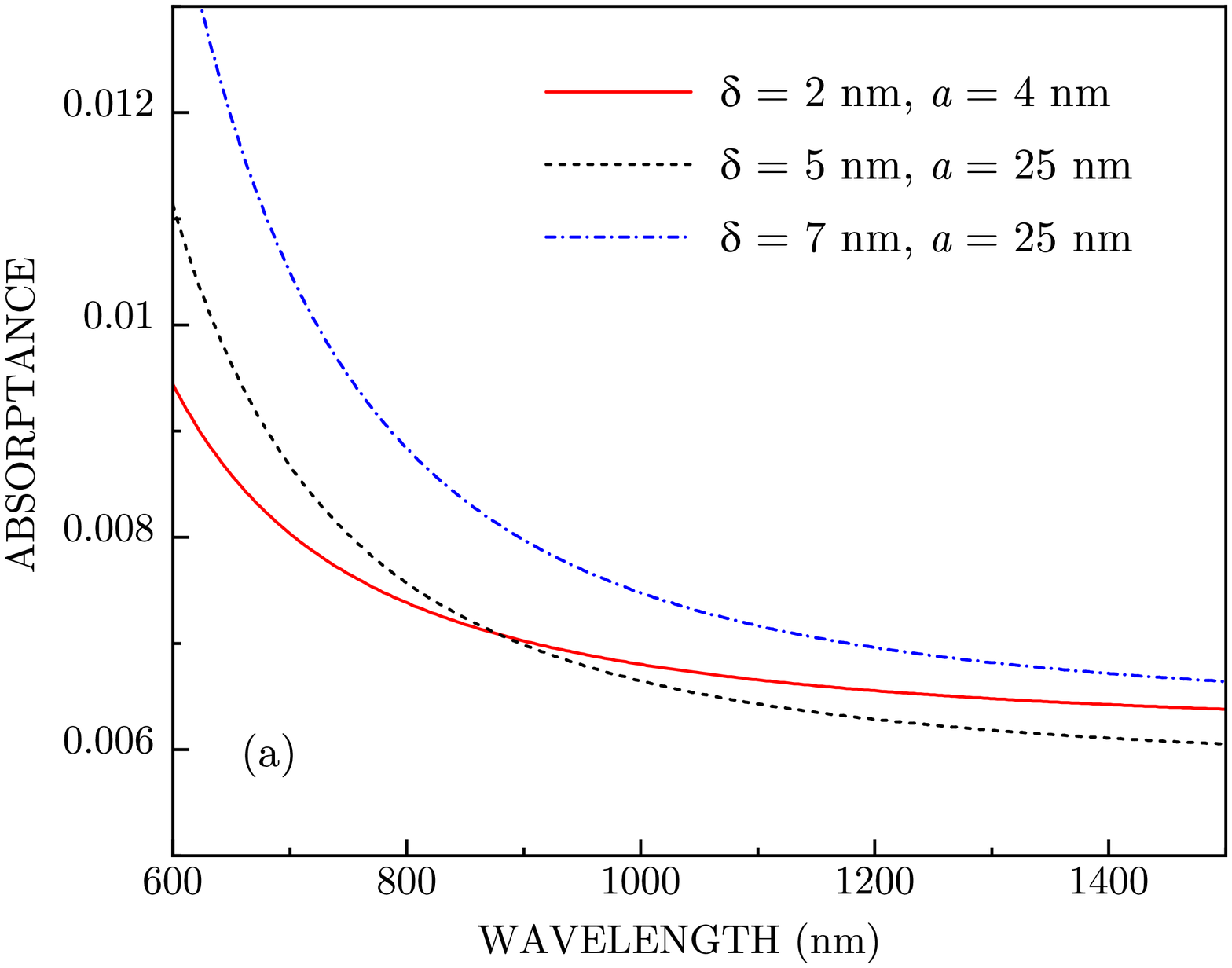}

\vspace{2mm}

\includegraphics[width=0.9\columnwidth]{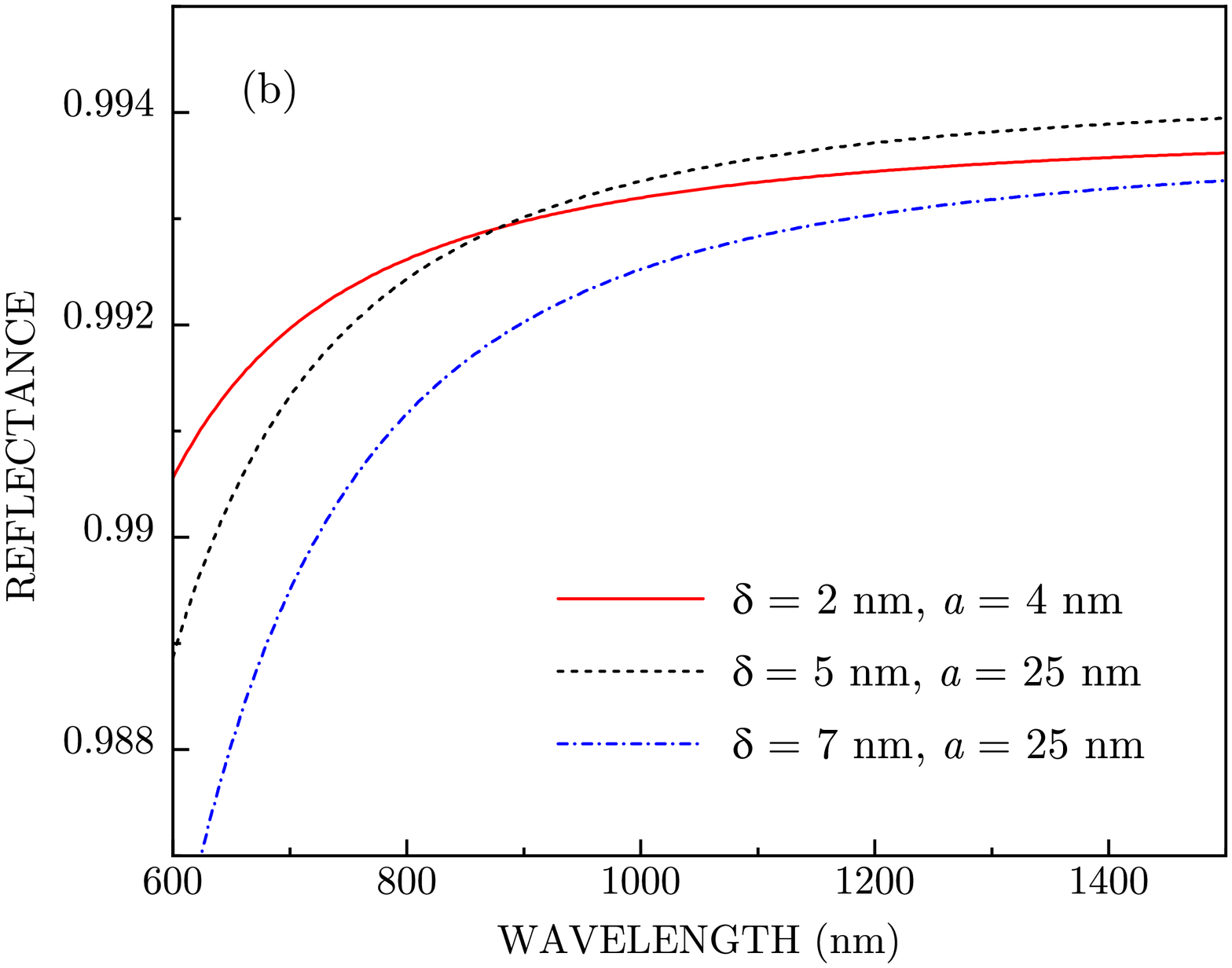}
\caption{\label{fig3} Calculated absorptance (a) and reflectance (b) at several values of roughness amplitude $\delta$ plotted against incident light wavelength  for different ratios $\delta/a$. In the long wavelength domain, the absorptance curves undergo reversal of order implying a larger absorption for smaller $\delta$.}
\end{center}
\vspace{-6mm}
\end{figure}
%

In Fig.~\ref{fig3}, we show the calculated absorptance and reflectance for several values of $\delta$ and different  ratios $\delta/a$. A striking feature in the long-wavelength spectral region, where the absorptance variations are primarily dominated by roughness effects,  is a \textit{larger} absorptance for roughness amplitude $\delta=2$ nm as compared to that for $\delta=5$ nm  but smaller ratio $\delta/a$ [see Fig.~\ref{fig3}(a)]. This surprising behavior can be understood, using the small-scale roughness asymptotics (\ref{finabs}), in terms of  competition between the second and third terms: the absorptance for parameter ratios $\delta/a=0.5$ and $\delta/d=0.08$ (solid curve) is larger than that for $\delta/a=0.2$ and $\delta/d=0.2$ (dashed curve) even though, in the latter case, the roughness amplitude $\delta$ is greater. At the same time, for shorter wavelengths, the absorptance curve for $a=25$ nm increases faster, as discussed above, and overtakes the $a=4$ nm absorptance curve at $\lambda\approx 870$ nm. Note that even for the ratio $\delta/a=0.5$, all roughness-induced corrections in Eq.~(\ref{finabs}) are still small. A similar reversal of order takes place for calculated reflectance $R=1-A$, as shown in Fig. \ref{fig3}(b). Such a reversal of order was reported in the absorption experiment \cite{springer04}, albeit for larger roughness.

%
\begin{figure}[tb]
\begin{center}
\includegraphics[width=0.9\columnwidth]{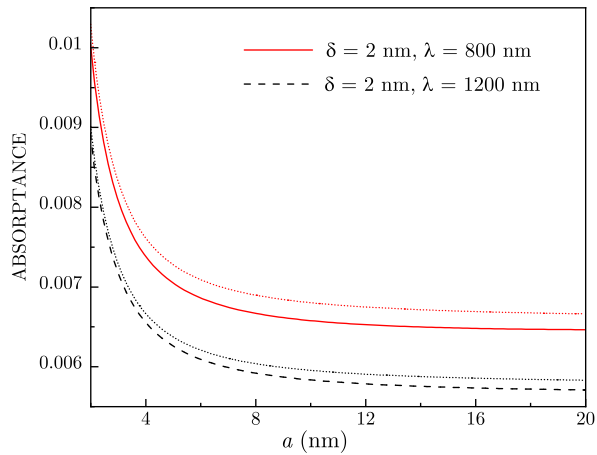}
\caption{\label{fig4} Calculated absorptance  at a small value of roughness amplitude $\delta=2$ nm plotted against the correlation length $a$ for incident light wavelengths $\lambda=800$ to 1200 nm. Dotted lines represent the small-scale roughness asymptotics calculated using Eq.~(\ref{finabs}).}
\end{center}
\vspace{-6mm}
\end{figure}
%

The above results indicate that, from the absorption perspective, the relevant parameter characterizing small-scale roughness is the ratio $\delta/a$ rather than the magnitude of $\delta$ relative to skin depth $d$ or wavelength $\lambda$. As discussed in Sec.~\ref{sec-ref}, this parameter characterizes increase in effective volume of the surface layer, in which the incident light is predominantly absorbed,  but it does not appear, to the best of our knowledge, in direct calculations of the reflection coefficient for rough metal surfaces \cite{Navarette09}. Note that the reflectance  evaluated as $R=1-A$, shown in Fig.~\ref{fig3}, represents the  total reflection, including the nonspecular one, which is difficult to evaluate accurately for the weak roughness case. At the same time, in the absorptance calculation, the parameter $\delta/a$ appears very naturally and, in fact, provides the dominant contribution for small-scale roughness. We stress that this contribution describes a "geometric" effect, as illustrated in Fig.~\ref{fig1}, and, therefore, it persists for any wavelength. To illustrate this point, in Fig.~\ref{fig4} we plot the absorptance dependence on correlation length $a$ for a small roughness amplitude $\delta=2$ nm and two wavelength values $\lambda=800$ to 1200 nm. For comparison, we also plot, by dotted lines, the corresponding asymptotics (\ref{finabs}) which show reasonably good agreements, especially for small $a$. While for $a>10$ nm ($\delta/a<0.2$) the absorptance changes only weakly, for smaller $a$ it sharply rises  in both cases, increasing by about 50\% for $a\sim\delta$.

%
\begin{figure}[tb]
\begin{center}
\includegraphics[width=0.9\columnwidth]{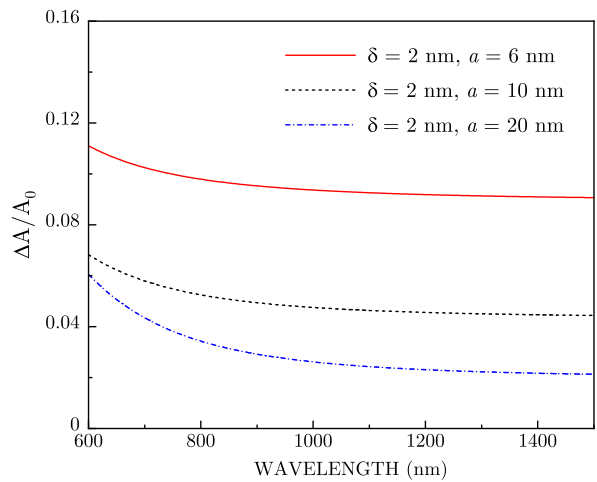}
\caption{\label{fig5} Calculated absorptance  fluctuations at a small value of roughness amplitude $\delta=2$ nm  shown for several values of correlation length $a$.}
\end{center}
\vspace{-6mm}
\end{figure}
%

The high sensitivity of the absorptance $A$ to roughness parameters implies that its relative variation $\Delta A/A_{0}$ can fluctuate substantially between samples characterized by  similar roughness amplitudes $\delta$ but larger variations of the correlation length $a$. In fact, such fluctuations can appear as variations of the Drude scattering rate $\gamma$, which were reported in high-precision ellipsometry measurements of the silver dielectric function \cite{yang15}. Indeed, in the Drude regime, we have $\varepsilon''\approx \omega_{p}^{2}\gamma/\omega^{3}$, $\kappa\approx \omega_{p}/\omega$, and $|t|^{2}\approx 4/\kappa^{2}$, so  that the smooth-surface absorptance can be estimated as $A_{0}=\varepsilon'' |t|^{2}/2\kappa\approx 2\gamma/\omega_{p}$. In this regime, we can present the rough-surface absorptance as $A=2\gamma_{\rm eff}/\omega_{p}$, where the effective scattering rate $\gamma_{\rm eff}$ has the form $\gamma_{\rm eff}=\gamma (1+\Delta A/A_{0})$. We then find that the apparent fluctuations of Drude scattering rate $\Delta \gamma =\gamma_{\rm eff}-\gamma$ are simply given by those of absorptance, i.e.,
\begin{equation}
\frac{\Delta \gamma}{\gamma}=\frac{\Delta A}{A_{0}}\approx \frac{\delta^{2}}{2a^{2}}+\frac{2\delta^{2}}{\sqrt{2\pi}da}+\frac{2\delta^{2}}{d^{2}},
\end{equation}
where, in the Drude regime, we used the asymptotic expression  (\ref{finabs}). In Fig. (\ref{fig5}), we plot the wavelength dependence of $\Delta A/A_{0}$ calculated for small-roughness amplitude $\delta=2$ nm and several values of $a$. In the long-wavelength spectral region $\lambda> 1000$ nm, the absorptance variations are nearly constant and could indeed appear as   fluctuations of $\gamma$, even though they are not caused by any electron scattering processes in metals. For the  parameters chosen,  such fluctuations can reach up to  10\% depending on the ratio $\delta/a$, which is comparable to the reported experimental values \cite{yang15}. Note finally that the increase of $\Delta A/A_{0}$ for shorter wavelengths, which was discussed above, here appears as "non-Drude" behavior of $\gamma_{\rm eff}$ that can be more or less pronounced for samples with different roughness parameters,  also consistent with the reported behavior of $\varepsilon''$ in this frequency domain \cite{yang15}.

\section{Conclusions}
\label{sec-conc}

In summary, we developed a perturbative approach for absorption of light in weakly rough opaque metal films characterized by a Gaussian surface profile with rms amplitude $\delta$ and correlation length $a$ that are smaller than  skin depth $d$ in the metal. We have shown that, in such systems,  the accurate choice of boundary conditions for unperturbed fields allows one to obtain the first-order roughness corrections to the absorptance  $A$ which otherwise would be excessively large.  We demonstrated that roughness-related absorptance variations $\Delta A$ are determined by the interplay between $a$ and $d$ which, in particular, can result in a larger absorption for smaller roughness amplitudes. We found that, for a small-scale roughness ($a\ll d$), the dominant contribution to $\Delta A$ comes from roughness-related increase in effective volume of the surface layer, in which  the incident light is predominantly absorbed, rather than from  light scattering from surface imperfections.  Accordingly, $\Delta A/A_{0}\sim \delta^{2}/a^{2}$ is nearly independent of the incident light wavelength and can fluctuate substantially between different samples characterized by small rms amplitudes $\delta$ but larger spread in $a$. We argued that such fluctuations, while not related to any electron scattering processes, can appear as  sample-to-sample variations of the Drude scattering rate  which could explain   uncertainties in the  imaginary part of the metal dielectric function reported in recent high-precision ellipsometry measurements \cite{yang15}.

Although we considered here the simplest case of normal incidence and one-dimensional roughness profile, it is straightforward to extend our approach to any incidence angle and polarization or to a two-dimensional roughness profile.  The different incidence angles would mainly affect the Fresnel coefficient $t$ that defines the smooth-surface absorptance $A_{0}$, while for $\kappa\gg 1$, the skin depth $d$ in the metal changes only weakly. For two-dimensional roughness, characterized by surface profile function $h(x,y)$, the effective volume of the surface layer,  in which  the incident light is predominantly absorbed, increases relative to that in the one-dimensional case.  In fact, the first term in Eq.~(\ref{finabs}) now doubles in magnitude, which implies more pronounced sample-to-sample apparent fluctuations of the Drude scattering rates \cite{yang15}. To the best of our knowledge, such fluctuations have not been previously described in direct calculations of the reflectance coefficient from rough metal surfaces, but here they emerge naturally when evaluating the absorptance. Note, finally, that the approach developed in this paper can also be used for  other roughness-related problems \cite{Gevorkian10,Gevorkian11}, as well as for describing absorption in lossy dielectric materials  \cite{Qin22}.

\section{Acknowledgments} This work was supported by  National Science Foundation  Grants  No. DMR-2000170, No. DMR-1856515,  and No.  DMR-1826886. Z.S.G. acknowledges support from Armenia Science Committee Grants No. 20RF-023 and  No. 21AG-1C062.

\end{document}